\documentclass[twocolumn]{aastex62} %{amsart}

\usepackage{xspace}
\usepackage{color}
\usepackage{rotating}
\usepackage{amsmath}
\usepackage{subfigure}
\usepackage[ampersand]{easylist}
\usepackage[version=4]{mhchem} % Molecular species! e.g. \ce{H2O}
\usepackage[normalem]{ulem}    % for strikeout text use: \sout{removed text}

\PassOptionsToPackage{hyphens}{url}\usepackage{hyperref}

\begin{document}

%% ApJ style
\bibliographystyle{apj}

%% Slugcomment
%\slugcomment{To be submitted by September 30, 2019}

\title{A mirage of the cosmic shoreline: \\ Venus-like clouds as a statistical false positive for exoplanet atmospheric erosion}

%% Short title, authors
\shorttitle{A Mirage of the Cosmic Shoreline}
\shortauthors{Lustig-Yaeger, Meadows \& Lincowski}

\correspondingauthor{Jacob Lustig-Yaeger}
\email{jlustigy@uw.edu}

\author[0000-0002-0746-1980]{Jacob Lustig-Yaeger}
\affiliation{Department of Astronomy and Astrobiology Program, University of Washington, Box 351580, Seattle, Washington 98195, USA}
\affiliation{NASA NExSS Virtual Planetary Laboratory, Box 351580, University of Washington, Seattle, Washington 98195, USA}

\author[0000-0002-1386-1710]{Victoria S. Meadows}
\affiliation{Department of Astronomy and Astrobiology Program, University of Washington, Box 351580, Seattle, Washington 98195, USA}
\affiliation{NASA NExSS Virtual Planetary Laboratory, Box 351580, University of Washington, Seattle, Washington 98195, USA}

\author[0000-0003-0429-9487]{Andrew P. Lincowski}
\affiliation{Department of Astronomy and Astrobiology Program, University of Washington, Box 351580, Seattle, Washington 98195, USA}
\affiliation{NASA NExSS Virtual Planetary Laboratory, Box 351580, University of Washington, Seattle, Washington 98195, USA}

%% Abstract %%

\begin{abstract}

Near-term studies of Venus-like atmospheres with JWST promise to advance our knowledge of terrestrial planet evolution. 
However, the remote study of Venus in the Solar System and the ongoing efforts to characterize gaseous exoplanets both suggest that high altitude aerosols could limit observational studies of lower atmospheres, and potentially make it challenging to recognize exoplanets as ``Venus-like''. 
To support practical approaches for exo-Venus characterization with JWST, we use Venus-like atmospheric models with self-consistent cloud formation of the seven TRAPPIST-1 exoplanets to investigate the atmospheric depth that can be probed using both transmission and emission spectroscopy. 
We find that JWST/MIRI LRS secondary eclipse emission spectroscopy in the 6 $\mu$m opacity window could probe at least an order of magnitude deeper pressures than transmission spectroscopy, potentially allowing access to the subcloud atmosphere for the two hot innermost TRAPPIST-1 planets. 
In addition, we identify two confounding effects of sulfuric acid aerosols that may carry strong implications for the characterization of terrestrial exoplanets with transmission spectroscopy: (1) there exists an ambiguity between cloud-top and solid surface in producing the observed spectral continuum; and (2) the cloud-forming region drops in altitude with semi-major axis, causing an increase in the observable cloud-top pressure with decreasing stellar insolation. Taken together, these effects could produce a trend of thicker atmospheres observed at lower stellar insolation---a convincing false positive for atmospheric escape and an empirical ``cosmic shoreline''. 
However, developing observational and theoretical techniques to identify Venus-like exoplanets and discriminate them from stellar windswept worlds will enable advances in the emerging field of terrestrial comparative planetology. 

\end{abstract}

%%% Keywords %%%

\section{Introduction\label{sec:intro}}

% 1a. General context of the work
Venus-like exoplanets pose unique opportunities and challenges for the near-term characterization of terrestrial exoplanet atmospheres \citep{Arney2018b}. 
Exo-Venuses are key near-term observational targets due to the transit bias in favor of finding and characterizing planets at short orbital period (e.g. high transit probability, high transit frequency, high equilibrium temperature), particularly in the \textit{TESS} era \citep{Ostberg2019}. 
Planets at similar insolation and with similar bulk properties to Venus are also favorable laboratories to empirically test runaway greenhouse theory, identify the location of the inner edge of the habitable zone (HZ), and probe the impact of atmospheric escape on an ensemble of terrestrial planets \citep{Kane2014b}. %, as well as being another point to compare the exoplanet ensemble with our own solar system. 
Additionally, the comparative study of Venus and Venus-like exoplanets are mutually beneficial research avenues \citep{Arney2018b}. Within the exoplanet population, 
if exo-Venuses are found to be common, they would point to a common end-state of terrestrial exoplanet evolution that Venus exemplifies. However, if true Venus analogs are rare, that may point to a more specific origin for Venus. 
%exo-Venuses may be the archetypal end-state of terrestrial planet evolution whose high occurrence rate warns of the perils of planetary evolution and hints at a common origin for Venus herself; or, true Venus analogs may be rare and suggest a more specific origin for Venus. 
Within the solar system, future orbiters and descent probes could provide detailed, in situ measurements to help answer outstanding questions about evolutionary processes and the current state of Venus, which will provide crucial context for the population of exo-Venuses \citep[see recent white papers:][]{Kane2018, Kane2019, Wilson2019}. 
% 1b. Narrower research area and statement of its importance
However, these exciting opportunities are contingent upon our ability to properly recognize and accurately characterize a Venus-like exoplanet when we see one. 

% However, probing beneath Venus's optically thick, global sulfuric acid clouds and hazes has been and continues to be a notoriously difficult endeavor, making remote sensing observations of Venus and, therefore any Venus-like exoplanets, extremely insensitive to the presence and character of any lower atmosphere and surface conditions.
Remote sensing observations have been used to understand and probe beneath the optically thick and global sulfuric acid clouds and hazes, which 
%make remote observations of Venus, and any Venus-like exoplanets, extremely insensitive to the presence and character of the lower atmosphere and surface. 
extend from 48 to 90 km altitude, and obscure the lower atmosphere and surface of Venus at most wavelengths. 
Although clouds were suspected early on due to Venus's high albedo and UV markings \citep{Hunten1983book}, their composition was unknown until optical phase curves ruled out water clouds \citep{Arking1968, Hansen1971a}, multi-band polarization phase curves matched the real index of refraction for a concentrated solution of sulfuric acid \citep{Hansen1971b}, and NIR absorption features confirmed \ce{H2SO4} \citep{Pollack1974}. 
These clouds thoroughly obscure the hot lower atmosphere at visible wavelengths, but the first clue to the extremely hot nature of the surface environment was a radio brightness temperature measurement of ${\sim} 560$ K at 3.15 cm by \citet{Mayer1958}, which was later confirmed by spacecraft observations \citep[e.g.][]{Barath1963} and descent probes \citep[e.g.][]{Marov1973}. 
Despite these challenges, peering beneath the clouds into the hot lower atmosphere has been possible with spectroscopy targeting 
%thermal emission windows on the night side of Venus \citep[e.g.][]{Allen1984, Allen1987, Carlson1991, Crisp1991}, which enable remote studies of the lower atmosphere \citep[e.g.][]{deBergh1995, Meadows1996, Barstow2012, Arney2014}.  
near-infrared windows on the Venus night side through which thermal emission from below the clouds escapes \citep[e.g.][]{Allen1984, Allen1987, Carlson1991, Crisp1991}, enabling remote studies of the Venus lower atmosphere and surface \citep[e.g.][]{Drossart1993, deBergh1995, Meadows1996, Barstow2012, Arney2014}. 

% Expected challenges to exo-Venus characterization based on Venus remote sensing
Extending the lessons learned from Venus remote sensing to the characterization of potential exo-Venuses 
%should raise concerns about our ability to accurately characterize the sulfuric acid clouds and probe the atmosphere beneath them because the historically most informative Venus observations lack feasible exoplanet analogs. This of course includes orbiters and descent probes, but also radio brightness temperature measurements, and in the near-term, precise optical and polarization phase curves. 
may be challenging as the historically most informative Venus observations lack feasible exoplanet analogs, either because they were made from orbiters or descent probes, or used radio brightness or precise optical and polarization phase curves. 

% Expected challenges to exo-Venus characterization using JWST
After launch, the \textit{James Webb Space Telescope} (JWST) will likely be used to attempt characterization of Venus-like exoplanets \citep{Barstow2016, Morley2017, Lincowski2018}, but these observations may be limited by how transmission and emission spectra are both significantly impacted by Venus-like clouds \citep{Lustig-Yaeger2019}. It has been well established by theory and observation that transmission spectroscopy is sensitive to obscuration by high altitude aerosols \citep[e.g.][]{Fortney2005, Berta2012, Ehrenreich2014, Knutson2014, Kreidberg2014, Nikolov2015, Morley2013, Charnay2015b, Charnay2015a}. Additionally, although dayside thermal emission, which is sensitive to the cloud deck temperatures, may be observed for exoplanets via secondary eclipse, nightside NIR thermal emission windows, which are sensitive to the lower atmosphere, may be significantly more challenging to observe for exoplanets when the far brighter dayside portion is included in the disk average. 
Despite these challenges, modeling efforts in advance of JWST indicate that the presence of cloudy Venus-like atmospheres could be detected for all seven planets in the TRAPPIST-1 system using JWST transmission spectroscopy to identify \ce{CO2} absorption features in the thin atmosphere above the clouds \citep{Lustig-Yaeger2019}. 

%Despite these challenges, modeling efforts in advance of JWST indicate that the presence of cloudy Venus-like atmospheres may be detected or ruled out for a select group of Earth-size planets transiting late M dwarfs. These include all seven planets in the TRAPPIST-1 system using JWST transmission spectroscopy to identify molecular absorption features \citep{Lustig-Yaeger2019}. 
%Additionally, methods have been proposed to efficiently detect atmospheres using JWST/MIRI secondary eclipse photometry to leverage a low dayside temperature due to the redistribution of heat to the nightside through the atmosphere \citep{Koll2019} or a high bond albedo due to subsolar clouds \citep{Mansfield2019}. Although promising for warm to hot planets ($T_{eq} = 300-880$ K), constraints from secondary eclipses become quickly infeasible for cooler planets into and beyond the HZ \citep{Lustig-Yaeger2019, Koll2019}. 

% 2a. Identification of a gap or other need for research
%Numerous modeling efforts have focused on exo-Venus atmospheres and how they can be characterized with future telescopes. 
%However, the extent of information that can be learned about the lower atmospheres, and the implications of this have not been discussed. 

% 3a. Summary of approach to answer the research question

% 3b. Announcement of principal findings and structure

% 4. Structure of paper

Another potential complication in identifying and interpreting spectra of Venus-like planets comes from the behavior of sulfuric acid cloud formation as a function of semi-major axis. 
%A string of Venuses, extending from interior to, through, and beyond the HZ could result following the super-luminous pre-main-sequence phase of late M dwarfs, like TRAPPIST-1 \citep{Lincowski2018}. 
The super-luminous pre-main-sequence phase of late M dwarfs, like TRAPPIST-1, could produce a string of Venuses, extending from interior to, through, and beyond the HZ \citep{Lincowski2018}. 
%During this time, each of the TRAPPIST-1 planets may have been subjected to runaway greenhouse water loss and subsequent oxygen build-up \citep{Luger2015}, followed by the loss or sequestration of oxygen and the outgassing of volatiles over time \citep{Schaefer2016, Garcia-Sage2017}, thus allowing high-\ce{CO2} Venus-like atmospheres to develop. 
During the pre-main-sequence phase, each of the TRAPPIST-1 planets may have been subjected to runaway greenhouse driven water loss and subsequent \ce{O2} buildup \citep{Luger2015, Bolmont2017, Lincowski2018}, even for planets well beyond the HZ. The subsequent sequestration of \ce{O2} and the outgassing of volatiles over time \citep{Schaefer2016, Garcia-Sage2017} may have allowed high-\ce{CO2} Venus-like atmospheres to develop. 

\citet{Lincowski2018} conducted a systematic study of the seven TRAPPIST-1 planets assuming they possess Venus-like atmospheres using a self-consistent 1D photochemical and climate model, which included sulfuric acid cloud formation. 
% Statement of fact from previous work -- The jumping off point 
Interestingly, these models demonstrated that sulfuric acid clouds form high in the atmospheres of hot Venus-like planets, but drop to lower altitudes for cooler Venus-like planets at lower incident stellar fluxes \citep{Lincowski2018}. 
% Observational implication and difficulty arising from the statement of fact
Since high altitude clouds can obscure molecular features in a transmission spectrum, the hottest cloudy exo-Venus atmospheres may actually be more difficult to detect than cooler cloudy exo-Venuses \citep{Lustig-Yaeger2019}---a practical manifestation of only probing the atmosphere above the clouds. 
% Corollary to the observable -- Initial presentation of the new problem
However, if only the upper, above-cloud, region of the atmosphere is readily probed, we may remain ignorant to the existence of a lower atmosphere, unable to distinguish cloud-top from solid surface.  
% Corollary to the new problem -- A potential misinterpretation from the new problem not being solved
%Furthermore, when surface and cloud top cannot be distinguished, any underlying physical (comparative planetology) trends in either surface pressure or cloud top pressure have the potential to be blended and misinterpreted through the degeneracy. 
%
%Since high altitude aerosols can obscure molecular features in a transmission spectrum by limiting access to the denser atmosphere below, the hotter cloudy exo-Venus atmospheres may actually be more difficult to detect than cooler cloudy exo-Venuses \citep{Lustig-Yaeger2019}, and they may mimic thinner atmospheres. 
%
Furthermore, the predicted increases in cloud top pressure with semi-major axis occur across a stellar insolation range that could also completely erode planetary atmospheres \citep{Dong2018}. 
Thus, observing a trend of thicker cloud-truncated atmospheres at lower stellar insolation may produce a statistical false positive for atmospheric escape across a population of terrestrial exoplanets \citep[e.g.][]{Bean2017, Checlair2019} and a mirage of the ``cosmic shoreline''---an empirical dividing line between planets with and without atmospheres \citep{Zahnle2017}. 

In this letter we explore two fundamental questions on the characterization of Venus-like exoplanets and their potential contribution to our understanding of terrestrial exoplanet atmospheric evolution:  
%\textit{will we recognize Venus-like exoplanets when we see them, and what might the consequences if we cannot?}
% Questions that this letter will address
\textit{(1) how do we infer the presence of and study sub-cloud atmospheres, and (2) what consequences and misinterpretations may arise if we cannot?}
% What will be tackled in this letter
In particular, we demonstrate how the presence of sulfuric acid clouds in \textit{thick} Venus-like atmospheres can mimic \textit{thin} cloud-free atmospheres in a transmission spectrum. We then explore how an observed decrease in cloud top altitude as a function of orbital distance could be misinterpreted as a surface pressure trend. Furthermore, such a trend with incident stellar flux could arise due to (1) atmospheric erosion via photoevaporation/thermal escape if the spectral continuum is assumed to be a solid surface, or (2) cloud top altitude variations due to condensation temperature if the continuum is assumed to be a cloud top. Finally, we offer observational and theoretical research avenues that may help to resolve this potential statistical false positive. 

In Section \ref{sec:methods} we describe the TRAPPIST-1 Venus-like atmospheric models used in this paper. In Section \ref{sec:results} we investigate the atmospheric regions probed by the transmission and emission spectra of Venus-like exoplanets applicable to JWST observations. In Section \ref{sec:discussion} we discuss the optimal paths towards inferring the presence of lower atmospheres for Venus-like exoplanets and we also expand on the hypothesis that, if we cannot detect lower atmospheres, atmospheric erosion could be invoked to explain mistakenly thin atmospheres, particularly in statistical characterization populations if Venus-like exoplanets are intrinsically common. We conclude in Section \ref{sec:conclusion}. 

\section{Methods} \label{sec:methods}

%\subsection{Models} \label{sec:methods:models}

We use the clear \ce{CO2} and cloudy Venus-like TRAPPIST-1 planet atmospheric models from \citet{Lincowski2018} as a foundation for the investigations in this paper. Briefly, \citet{Lincowski2018} used the VPL Climate model, a 1D radiative-convective equilibrium climate model applicable to terrestrial planet atmospheres \citep{Meadows2018, Robinson2018b}. The climate model is coupled to a 1D atmospheric photochemistry model originally developed by \citet{Kasting1979} and significantly improved upon by \citet{Zahnle2006}; this code is described in detail in \citet{Meadows2018} and has been used extensively for terrestrial exoplanet photochemical modeling across a broad range of redox states \citep[e.g.][]{Segura2005, Arney2016, Arney2017, Schwieterman2016, Arney2019}. In particular, the photochemical code was specifically updated and validated for modeling Venus-like atmospheres \citep{Lincowski2018}. 

%\citet{Lincowski2018} modeled Venus-like sulfuric acid aerosols self-consistently with the photochemical forcing of the late M dwarf SED and the radiative-convective climate model. 
\citet{Lincowski2018} used radiatively-active, photochemically-self-consistent sulfuric acid aerosols in their climate and spectral calculations. 
These calculations considered the photochemical production of \ce{H2SO4} vapor given the forcing from the late M dwarf SED, the temperature-dependent condensation of \ce{H2SO4}, the sedimentation of \ce{H2SO4} condensates, and their thermal decomposition at high temperatures in the lower atmosphere. Together these effects determined the aerosol effective radii and \ce{H2SO4} concentration in each layer. 
%Sulfuric acid aerosols modes were modeled similar to those described in \citet{Crisp1986} for Venus. 
%\citet{Lincowski2018} used refractive indices for sulfuric acid solutions from \citet{Palmer1975}, and calculated mass-conserving log-normal aerosol particle distributions, from which mie-scattering phase functions and optical depths were calculated. 
\citet{Lincowski2018} used refractive indices for sulfuric acid solutions from \citet{Palmer1975} 
ranging in concentration from 25-100\%, which were calculated for each atmosphere using the vapor pressure equilibrium between the \ce{H2O} and \ce{H2SO4} gases and the condensed \ce{H2SO4} solution.  
Mass-conserving log-normal aerosol particle distributions were computed (with geometric standard deviation equal to 0.25, as used in \citet{Crisp1986} for Venus), from which mie-scattering phase functions and optical depths were calculated. 
We note that \citet{Lincowski2018} found that sulfuric acid clouds did not form for a Venus-like TRAPPIST-1 b because the atmosphere was too hot for them to condense, so TRAPPIST-1 b is omitted from our cloudy Venus-like cases. We discuss the observational implications of this result in Section \ref{discussion:mitigation}. 

Transmission and emission spectra of the TRAPPIST-1 Venus-like planets were produced in \citet{Lincowski2018} using the Spectral Mapping Atmospheric Radiative Transfer (SMART) code \citep[developed by D. Crisp; ][]{Meadows1996}. SMART is a line-by-line, multi-stream, multi-scattering radiative transfer code that includes layer dependent gaseous and aerosol absorption and scattering, and treats both stellar and thermal source functions. 
SMART also calculates transmission spectra for transiting exoplanets using a ray tracing algorithm that includes the refraction of stellar light passing through the atmosphere \citep{Misra2014a, Robinson2018}.  
Gaseous rotational-vibrational line absorption coefficients were calculated using the LBLABC model \citep{Meadows1996} with the HITEMP2010 and HITRAN2012 line lists \citep{Rothman2010, Rothman2013}. 
%Venus-like sulfuric acid aerosols were modeled similar to the Venus aerosol modes described in \citet{Crisp1986}. \citet{Lincowski2018} used radiatively-active, photochemically-self-consistent sulfuric acid aerosols in the climate and spectral calculations.
%They used refractive indices for sulfuric acid solutions from \citet{Palmer1975}, and calculated mass-conserving log-normal aerosol particle distributions, from which mie-scattering phase functions and optical depths were calculated. 

To understand the maximum possible depth that can be probed into clear and cloudy Venus-like atmospheres, we processed the atmospheric structure and spectra of the Venus-like TRAPPIST-1 models from \citet{Lincowski2018} to reveal the average pressure into each atmosphere that can be probed with transmission and emission spectroscopy. 
For transmission spectroscopy, we used the relationship between altitude and pressure for atmospheres in hydrostatic equilibrium to interpolate the effective transit height to an effective transit pressure. Although this so called ``transit pressure'' is not a direct observable, it is tied to the observable transit depth, $(R_p/R_s)^2$, and approximates the depth into the atmosphere at which it becomes optically thick in the slant transit geometry. While this is similar to the effective transit height, the transit height increases radially out of the atmosphere and must assume a zero-point altitude that presumes knowledge of the planet's solid body radius---a key point of interest---but which is unknown \textit{a priori}. Alternatively, the transit pressure increases into the atmosphere from space---a known zero point pressure boundary condition---to the maximum pressure probed, and is therefore a good measure of how deep into the atmosphere the transmission spectrum probes. The maximum transit pressure across a given wavelength range also provides forward modeling insight into the cloud top pressure or reference pressure that would be inferred by atmospheric retrievals \citep[e.g.][]{Benneke2013, Kreidberg2015, Line2016, Benneke2015}. 

To assess the depth probed into Venus-like atmospheres with emission spectroscopy, we recomputed the radiative transfer with SMART for the \citet{Lincowski2018} Venus-like models to solve for the atmospheric pressure at which the total optical depth is unity at normal incidence. SMART calculates the pressure of optical depth unity in terms of Rayleigh scattering, gaseous absorption, and aerosol extinction. We combine these three terms into a single ``emission pressure'' or ``brightness pressure'' spectrum, which reflects the dominating process at each wavelength and approximates the pressure from which thermal emission emerges the atmosphere. 
Note that unlike the transit pressure, which resembles the observable transmission spectrum $(R_p/R_s)^2$, the ``emission pressure'' is significantly different from the observable eclipse depths of an emission spectrum ($F_p / F_s$), which 
%increase by roughly two orders of magnitude between about 5-15 $\mu$m for any single planet's flux relative to the star, and which decrease by roughly four orders of magnitude (at ${\sim} 7.5$ $\mu$m) from TRAPPIST-1 b to TRAPPIST-1 h because colder planets emit less thermal flux. 
have a strong wavelength dependence in accordance with the planet and star fluxes. Rather, the emission pressure more closely resembles the transit pressure, allowing for qualitative and quantitative comparisons between the two observational techniques.  

\section{Results} \label{sec:results}

\subsection{Transmission Spectroscopy}

\begin{figure*}[t]
\centering
\includegraphics[width=0.98\textwidth]{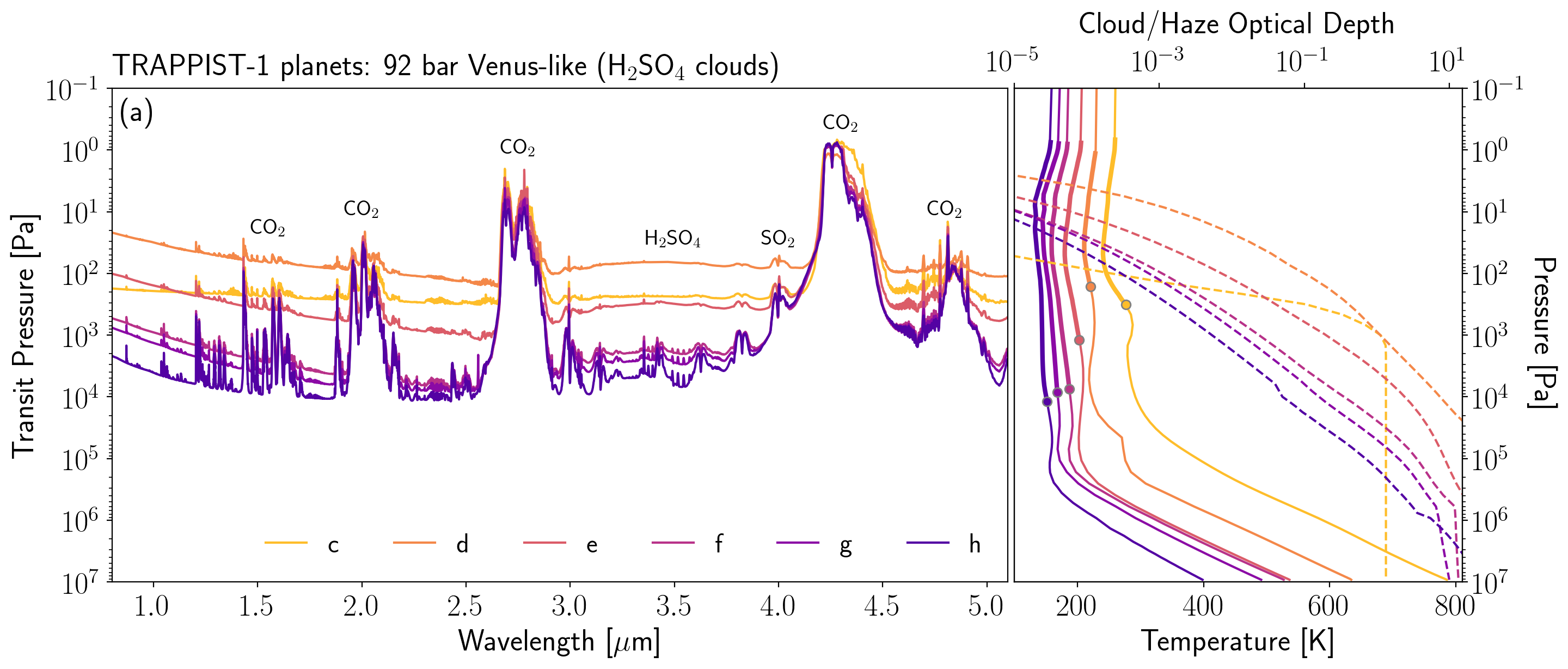}
\includegraphics[width=0.98\textwidth]{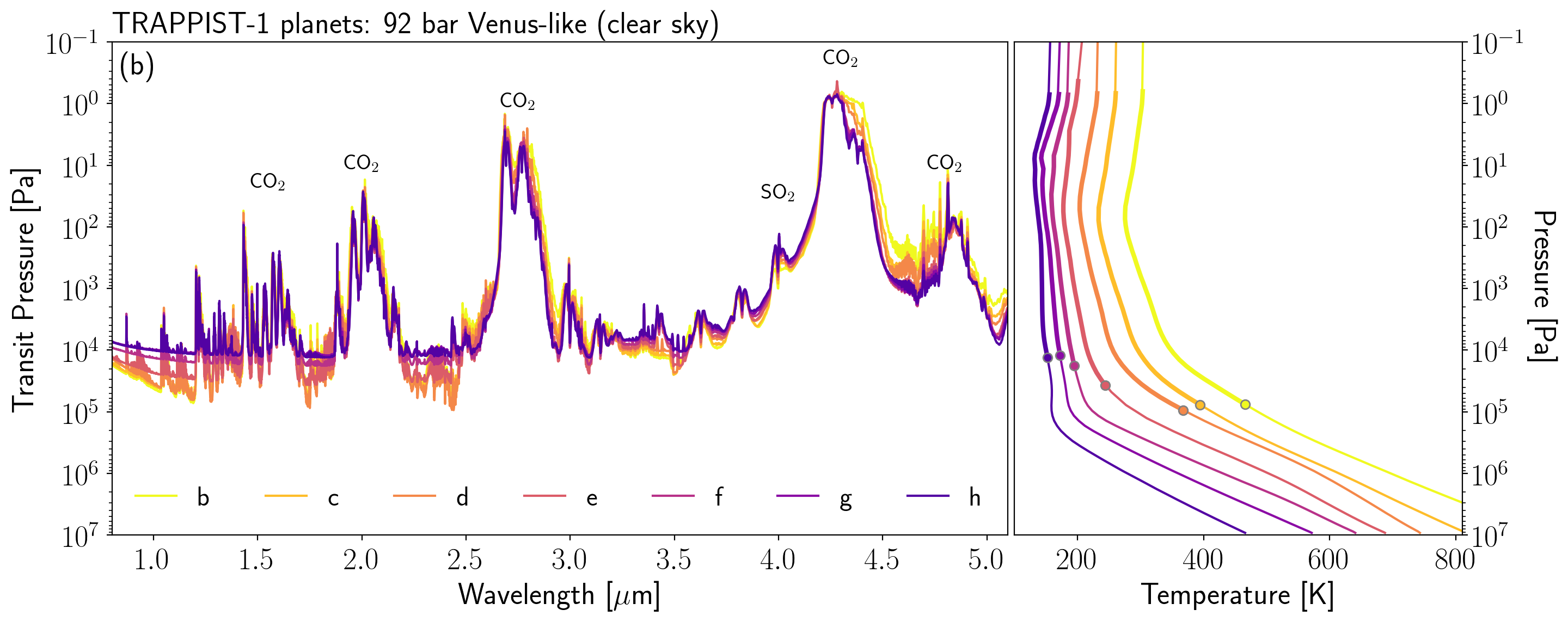}
\caption{Transmission spectrum models of the TRAPPIST-1 planets assuming they possess thick Venus-like atmospheres with \ce{H2SO4} clouds (top panels) and without clouds (bottom panels) \citep[models from][]{Lincowski2018}. The left panel shows transmission spectra in units of pressure probed into the atmosphere. The right panel shows the atmospheric temperature structure (solid lines; lower axis) and the cumulative vertical cloud/haze optical depth from the top of the atmosphere (dashed lines; upper axis). Thicker line styles indicate the region of the atmosphere that the transmission spectrum is sensitive to, and dots are shown on the temperature profiles to indicate the highest possible pressure that may be probed by the transmission spectrum for such planets. \textbf{\ce{H2SO4} clouds effectively prevent the lower atmosphere of Venus-like exoplanets from being remotely sensed by a transmission spectrum. However, for the cooler planets, cloud formation occurs at lower altitudes and enables the transmission spectrum to probe deeper into the atmosphere.}}
\label{fig:transit_pressure}
\end{figure*}

To explore the impact of clouds on the range of pressures probed by transmission spectroscopy, Figure \ref{fig:transit_pressure} shows model transmission spectra for each TRAPPIST-1 planet with and without \ce{H2SO4} clouds in the upper and lower panels, respectively. The transmission spectrum is shown between about 1-5 $\mu$m---the range of the JWST/NIRSpec Prism instrument, which is optimal for detecting the atmospheres of the TRAPPIST-1 planets \citep[][]{Batalha2018, Lustig-Yaeger2019}---in units of the pressure into the atmosphere that is probed. 
%Although this so called ``transit pressure'' is not a direct observable, it is intimately tied to the observable transit depth, $(R_p/R_s)^2$, and the ``transit altitude'', or altitude above the surface at which the atmosphere becomes optically thick in the slant transit geometry, which are both common units to report simulated transmission spectra. 
%Furthermore, transit altitude increases radially out of the atmosphere and must assume a zero-point altitude that amounts to knowledge of the planet's solid body radius, which is a key point of interest, but unknown \textit{a priori}, whereas transit pressure increases into the atmosphere from space to the maximum pressure probed. As a result, the transit pressure is a good measure of how deep into the atmosphere the transmission spectrum reaches. 
The right panels of Fig. \ref{fig:transit_pressure} show the thermal structure of each TRAPPIST-1 planet atmosphere on the same pressure y-axis as the transmission spectrum. Thicker line styles indicate the vertical region of the atmosphere that the transmission spectrum is sensitive to, and the circular points denote the maximum atmospheric pressure that is probed. The top-right panel also shows the cumulative vertical optical depth for the \ce{H2SO4} aerosols on the top x-axis, which increases going down into the atmosphere. 

Although the clear and cloudy transmission spectra appear similar due to the common presence of \ce{CO2} absorption bands, their respective continua vary by up to 3 orders of magnitude in pressure, which affects the strength and detectability of \ce{CO2} absorption  \citep[see][]{Lustig-Yaeger2019} and the depth into the atmosphere that may be probed by the spectrum. 
In units of pressure, the transmission spectra of different planets with atmospheres of similar compositions look quite similar, despite having different radii, masses, and temperatures. 
At wavelengths where the atmosphere is optically thick due to strong \ce{CO2} absorption (e.g. 2.7, 4.3, and 15.0 $\mu$m), 
%all of the TRAPPIST-1 planets show absorption features in their transmission spectra at roughly the same low atmospheric pressure (1-10 Pa), irrespective of the presence of clouds (and the incident stellar flux, and planet mass and radius). That is, within the strong \ce{CO2} bands all of the planets are probed down to the same depth in the atmosphere. 
the opacity is sufficiently high above the clouds that the peak absorption in the bands occurs at the same pressure in the upper atmosphere (1-10 Pa) for all of the TRAPPIST-1 planets, uninfluenced by the clouds and hazes at higher pressures below. 
At wavelengths where the atmosphere is optically thin, the presence of \ce{H2SO4} aerosols significantly raises the spectral continuum altitude to lower atmospheric pressures. For instance, the continuum pressure at 2.5 $\mu$m is $10^5$ Pa and $10^2$ Pa for clear and cloudy TRAPPIST-1d models, respectively. These cloudy results show both a significant departure from the clear atmosphere cases for each TRAPPIST-1 planet (${>} 100 {\times}$ lower continuum pressures for the cloudy inner planets compared to clearsky) and a significant variance in the pressure of the spectral continuum from one cloudy planet to the next (${\sim} 100 {\times}$ lower continuum pressure for the cloudy inner planets compared to the cloudy outer planets).   

\begin{figure*}[t]
\centering
\includegraphics[width=0.48\textwidth]{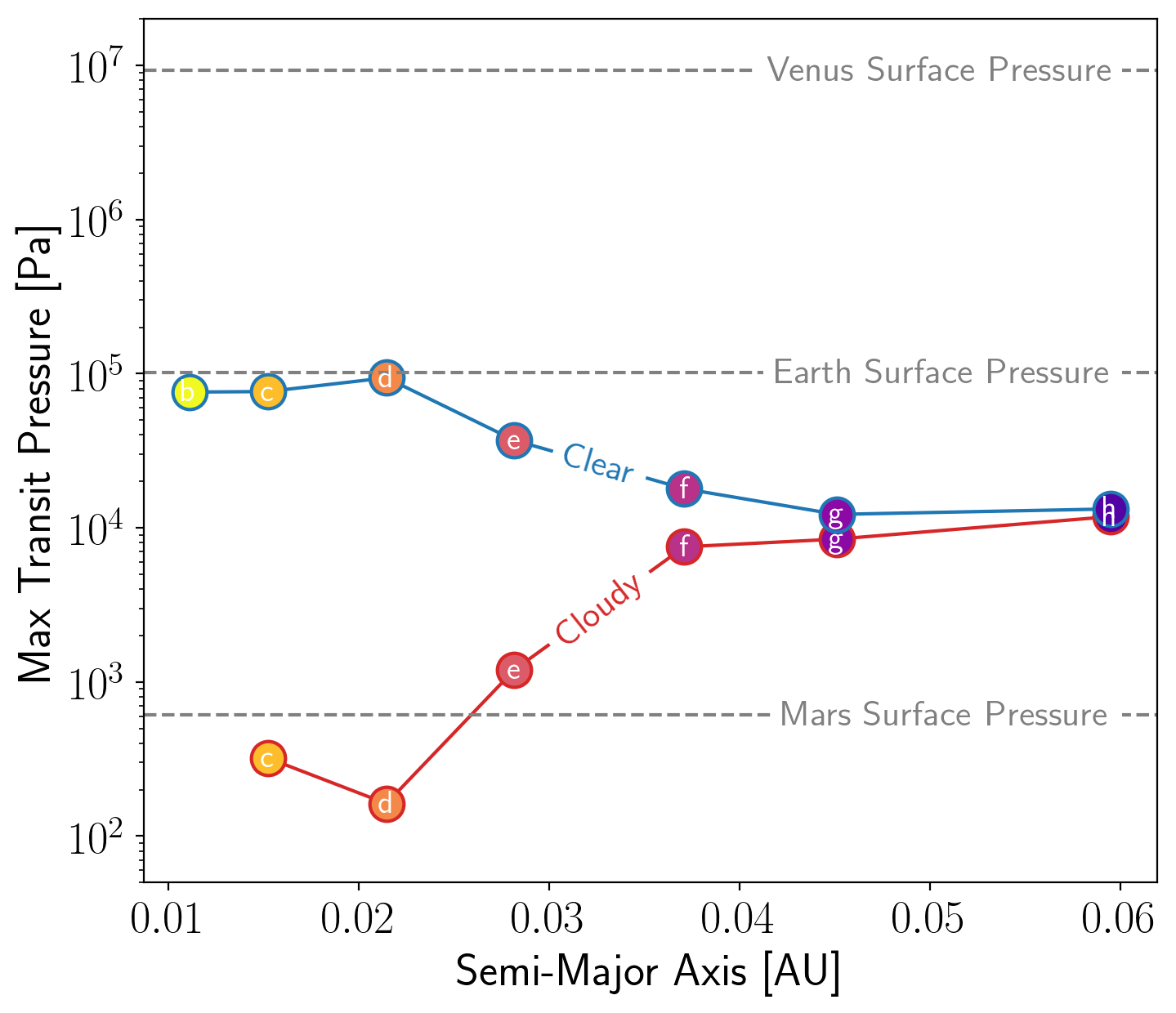}
\includegraphics[width=0.48\textwidth]{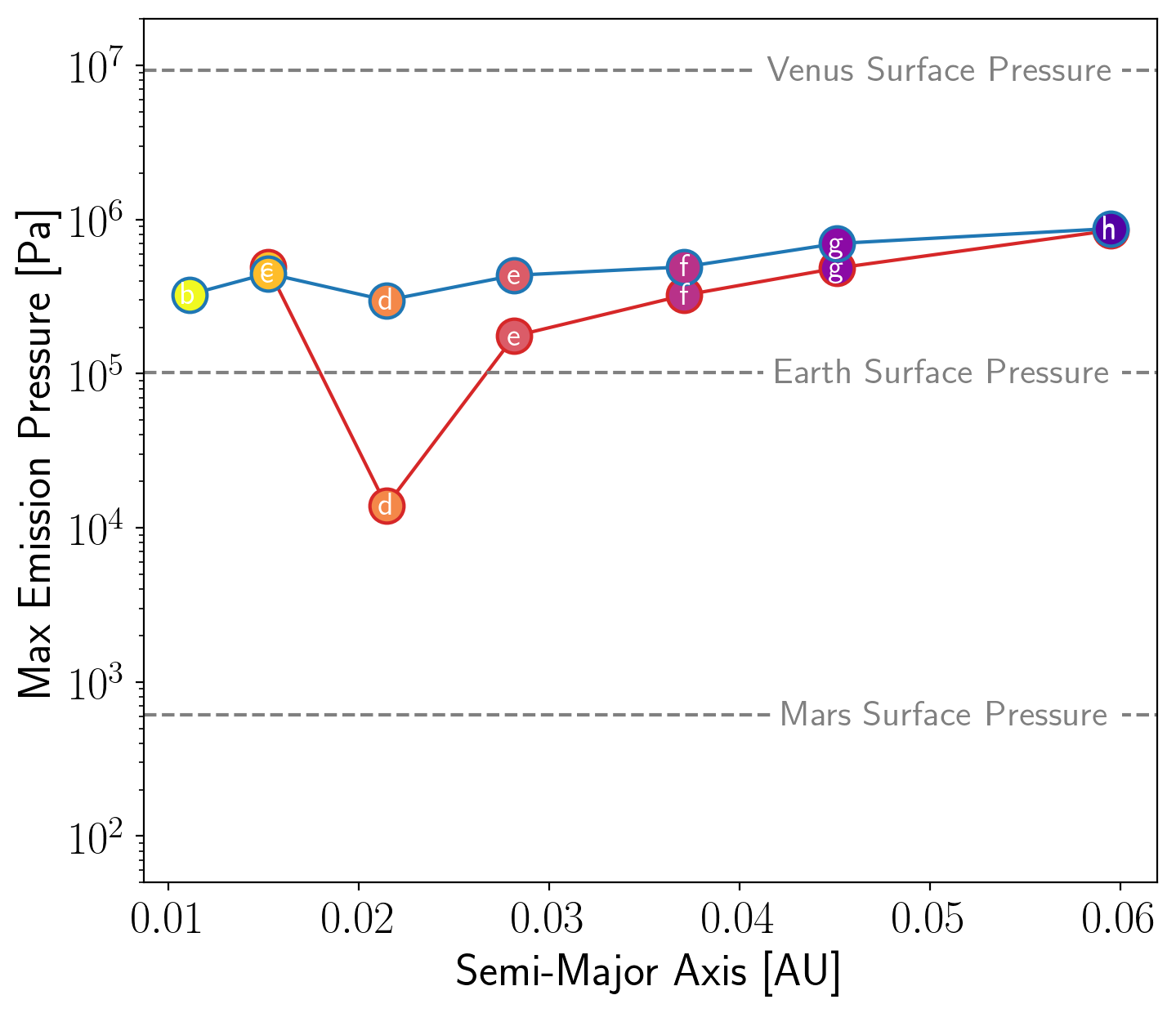}
\caption{Maximum pressure probed in the transmission spectrum (left panel) and emission spectrum (right panel) of clear (blue line) and cloudy (red line) Venus-like models of the TRAPPIST-1 planets as a function of semi-major axis. The surface pressures of Venus, Earth, and Mars are shown as dashed horizontal grey lines for reference. \textbf{The maximum transit pressure appears to increase with semi-major axis for models with \ce{H2SO4} clouds, while it decreases with semi-major axis for clear sky models. Although the maximum pressure probed in emission exceeds that of transmission, it does not approach the simulated Venusian surface pressure. }}
\label{fig:transit_probed}
\end{figure*}

The left panel of Figure \ref{fig:transit_probed} shows the maximum pressure probed by the transmission spectra (shown in Fig. \ref{fig:transit_pressure}) as a function of semi-major axes for the TRAPPIST-1 planets. Clear Venus-like atmospheres (blue lines) have their highest pressures accessible in the 1-3 $\mu$m range between the \ce{CO2} bands (e.g. at about 1.7, 2.4, and 3.1 $\mu$m), while Venus-like atmospheres with \ce{H2SO4} clouds (red lines) have their highest pressures accessible in the 2.4-2.6 $\mu$m range where \ce{H2SO4} aerosol scattering is weakest and just short of the 2.7 $\mu$m \ce{CO2} band, which has notably pressure broadened wings when not obscured by aerosols. These wavelengths offer the best opportunity to probe deepest into Venus-like atmospheres in transmission, and offer observation leverage for retrieving a cloud-top or surface pressure. 

There are no wavelengths at which the transmission spectra of Venus-like TRAPPIST-1 planets access their lower atmospheres. For the inner TRAPPIST-1 planets, if they have Venus-like atmospheres, then the transmission spectrum will only probe down to about the Martian surface pressure (610 Pa). If they are not cloudy, then they may be probed down to about the surface pressure of Earth (101 kPa). For the outer TRAPPIST-1 planets, the presence of clouds minimally affects the maximum transit pressure, and yet they still cannot be probed to higher pressures than about $10^{4}$ Pa. At least two, and up to five, orders of magnitude in pressure exist between the maximum transit pressure and the unseen Venusian surface pressure, which these models share. 

Despite the inaccessibility of lower atmospheres, clear and cloudy atmospheres exhibit distinctly opposing trends in the depth into their atmospheres that may be probed as a function of semi-major axis. Clear \ce{CO2} atmospheres gently slope from higher pressures accessible for the inner planets to lower pressures for the outer planets. However, Venus-like atmospheres with \ce{H2SO4} clouds generally increase from lower pressures accessible for the inner planets to higher pressures accessible for the outer planets. 

The divergent scaling with semi-major axis seen between clear and cloudy atmospheres is a result of the underlying physics that controls the transmission spectrum continuum. 
For clear atmospheres, refraction places a fundamental limit on the depth into the atmosphere that can be accessed by a transmission spectrum \citep{Betremieux2014, Misra2014a}. Clear atmospheres with larger semi-major axes cannot be probed as deeply as those closer to the star due to the dependence of the critical refraction pressure on (1) the angular size of the host star as seen from the transiting planet and (2) the scale height of the planetary atmosphere, both of which decrease with semi-major axis for similar composition planets within the same planetary system.  

For cloudy atmospheres, the optically thick cloud deck limits the depth into the atmosphere that can be probed by the transmission spectrum. 
%However, since cloud formation processes tend to favor lower altitude clouds for planets with larger semi-major axes, \citet{Lincowski2018} demonstrate a steady transition from high altitude clouds at low pressures for the inner TRAPPIST-1 planets to low altitude clouds at higher pressures for the outer TRAPPIST-1 planets. 
However, since the fixed temperature of sulfuric acid cloud condensation occurs lower in the atmospheres (at higher pressures) for cooler planets,  \citet{Lincowski2018} demonstrate a steady transition from high altitude clouds at low pressures for the inner TRAPPIST-1 planets to low altitude clouds at higher pressures for the outer TRAPPIST-1 planets.
This cloud formation trend is manifested in the observable transmission spectrum continuum, and enables increasingly higher pressures to be accessed for planets with increasing distance from their parent star. 
The clear and cloudy trends in Figure \ref{fig:transit_probed} converge for planets at sufficiently large orbital separations (e.g. TRAPPIST-1 f, g and h) as the cloud tops drop below the critical refraction pressure. 

\begin{figure*}[t]
\centering
\includegraphics[width=0.98\textwidth]{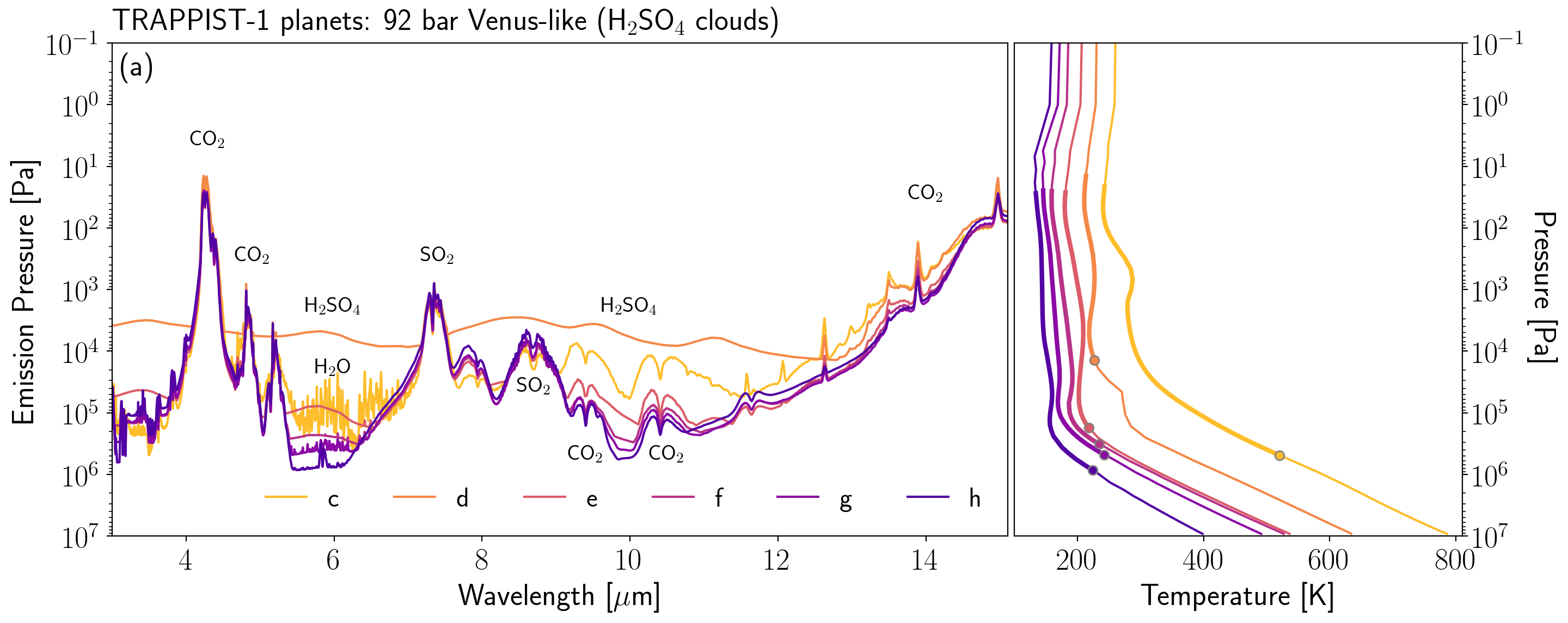}
\includegraphics[width=0.98\textwidth]{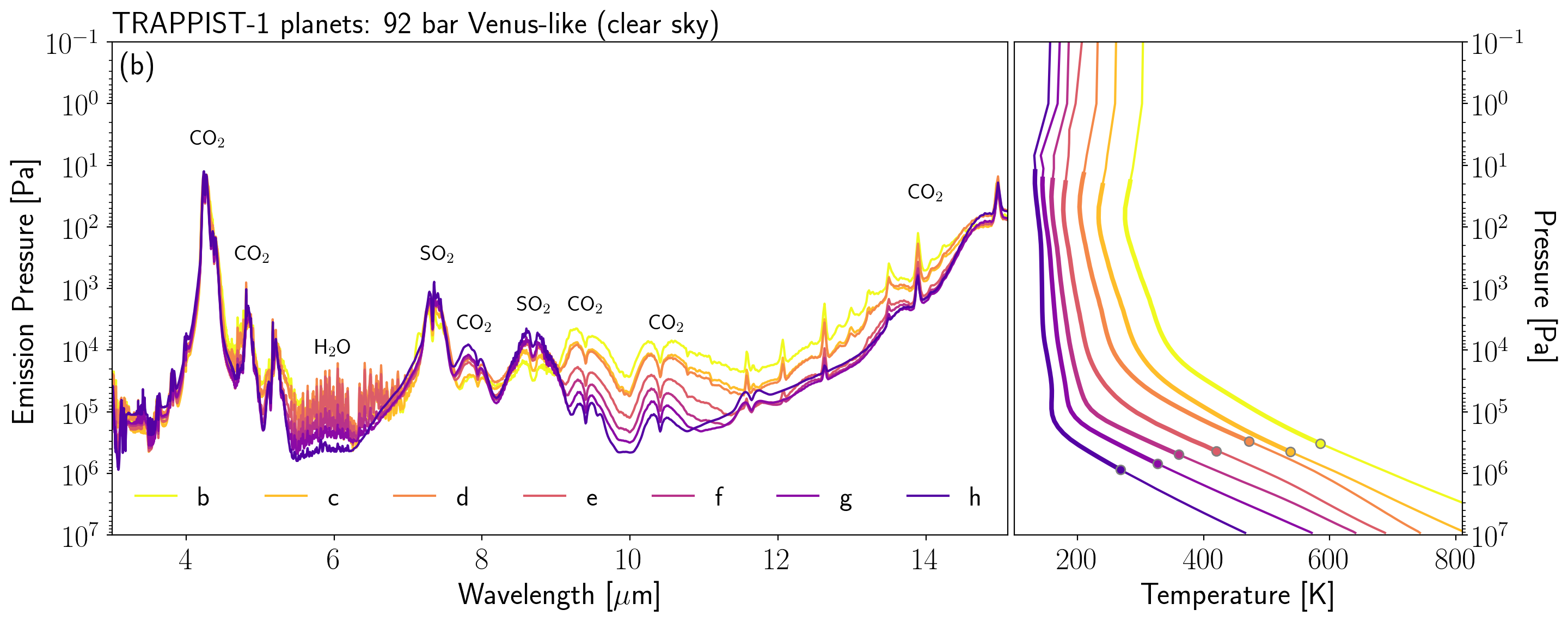}
\caption{Analogous to Figure \ref{fig:transit_pressure} but for emission spectrum models of the TRAPPIST-1 planets assuming they possess thick Venus-like atmospheres with \ce{H2SO4} clouds (top panels) and without clouds (bottom panels). The left panel shows the atmospheric pressure from which the planets thermal emission emanates ($\tau = 1$). \textbf{Various NIR and MIR atmospheric windows, which enable the planet to thermally cool to space, may allow distant observers to detect the higher pressures and temperatures of the sub-cloud atmosphere, but these windows can be closed by aerosol extinction (e.g. TRAPPIST-1 d) and gaseous absorption (e.g. \ce{H2O} at 6 $\mu$m), but they are infeasible for JWST to detect for planets cooler than TRAPPIST-1 c.}}
\label{fig:emission_pressure}
\end{figure*}

\subsection{Emission Spectroscopy} 

%We now turn to emission spectroscopy to compare and contrast accessible atmospheric regions with transmission spectroscopy. 
We also consider the limits of atmospheric study for Venus-like planets using emission spectroscopy. 
Figure \ref{fig:emission_pressure} shows the depth into the atmosphere that is probed by a secondary eclipse thermal emission spectrum for Venus-like models of the TRAPPIST-1 planets with and without \ce{H2SO4} clouds. Similar to Figure \ref{fig:transit_pressure}, the left panels of Figure \ref{fig:emission_pressure} show the ``emission pressure'', or the average pressure level in the atmosphere from which most of the thermal emission emerges at normal incidence. We show the emission pressure over a wavelength range that it is applicable to JWST's Mid-IR Instrument (MIRI) Low Resolution Spectrometer (LRS), which is optimal for observing thermal emission from the TRAPPIST-1 and similar exoplanets during secondary eclipse \citep{Lustig-Yaeger2019}. The right panels of Figure \ref{fig:emission_pressure} show the atmospheric thermal structure on the same pressure y-axis as the emission pressure for comparison, with line thickness highlighting the thermally emitting region of the atmosphere. 

%Note that unlike the ``transit pressure'' shown in Fig. \ref{fig:transit_pressure}, which strongly resembles the observable transmission spectrum, the ``emission pressure'' is significantly different from the observable eclipse depths of an emission spectrum ($F_p / F_s$), which increase by roughly two orders of magnitude between about 5-15 $\mu$m for any single planet's flux relative to the star, and which decrease by roughly four orders of magnitude (at ${\sim} 7.5$ $\mu$m) from TRAPPIST-1 b to TRAPPIST-1 h because colder planets emit less thermal flux. 

Numerous transparent windows in the near- and mid-IR offer glimpses into the deeper atmosphere of Venus-like planets. For instance, at 6 $\mu$m there is a prominent window where thermal emission can be seen coming from pressures of about $10^5 - 10^6$ Pa. There are also windows in the NIR that probe even deeper into the atmosphere, but which are not shown here due to the insensitivity of JWST to thermal emission in the NIR. Note, however, that these NIR windows have been used extensively to study the surface and near-surface of Venus \citep[see][]{Meadows1996, deBergh2006}.  

The right panel of Figure \ref{fig:transit_probed} shows the maximum pressure reached by an emission spectrum over the MIRI LRS bandpass as a function of semi-major axis for the TRAPPIST-1 planets. 
In all cases considered here, emission spectroscopy probes higher pressures than transmission spectroscopy. 
In general, cloudy atmospheres emit from higher altitudes and lower pressures than clear atmospheres. 
However, even thermal emission from the clear atmospheres is coming from over an order of magnitude lower pressures than the surface. 

Whereas transmission spectroscopy is more sensitive to the location of the cloud top due to the slant optical depth \citep{Fortney2005}, emission spectroscopy is more sensitive to the total optical depth of the clouds in the atmospheric column. The maximum emission pressure trend with semi-major axis for the cloudy Venus-like models notably tracks the total extinction optical depth of the aerosols, seen in Figure 7\footnote{\href{http://www.astroexplorer.org/details/apjaae36af7}{Link to figure on the Astronomy Image Explorer}} of \citet{Lincowski2018}. 
TRAPPIST-1 c and d effectively bracket the small particle haze and thick cloud regimes, respectively, which both exist in the Venus atmosphere \citep{Crisp1986}. 
That is, TRAPPIST-1 c's total \ce{H2SO4} aerosol extinction optical depth is of order unity, which is does not substantially modify the thermal emission spectrum from the clear sky case. However, TRAPPIST-1 d's \ce{H2SO4} aerosol total optical depth peaks among the TRAPPIST-1 planets at $\tau {\sim} 30$, due to the strong formation rate, cooler temperature, and lower gravity, which allows larger particles to be sustained and suspended. The resulting extended haze and cloud layer significantly mutes spectral features in the thermal emission spectrum and restricts remote sensitivity to the lower atmosphere. Beyond TRAPPIST-1 d, the total aerosol optical depth decreases with semi-major axis, revealing higher atmospheric pressures and explaining the convergence of the clear and cloudy lines in the right panel of Figure \ref{fig:transit_probed}. 
However, observing thermal emission spectra from cool terrestrial exoplanets is not feasible with JWST, which we discuss next. 

\begin{figure*}[t]
\centering
\includegraphics[width=0.98\textwidth]{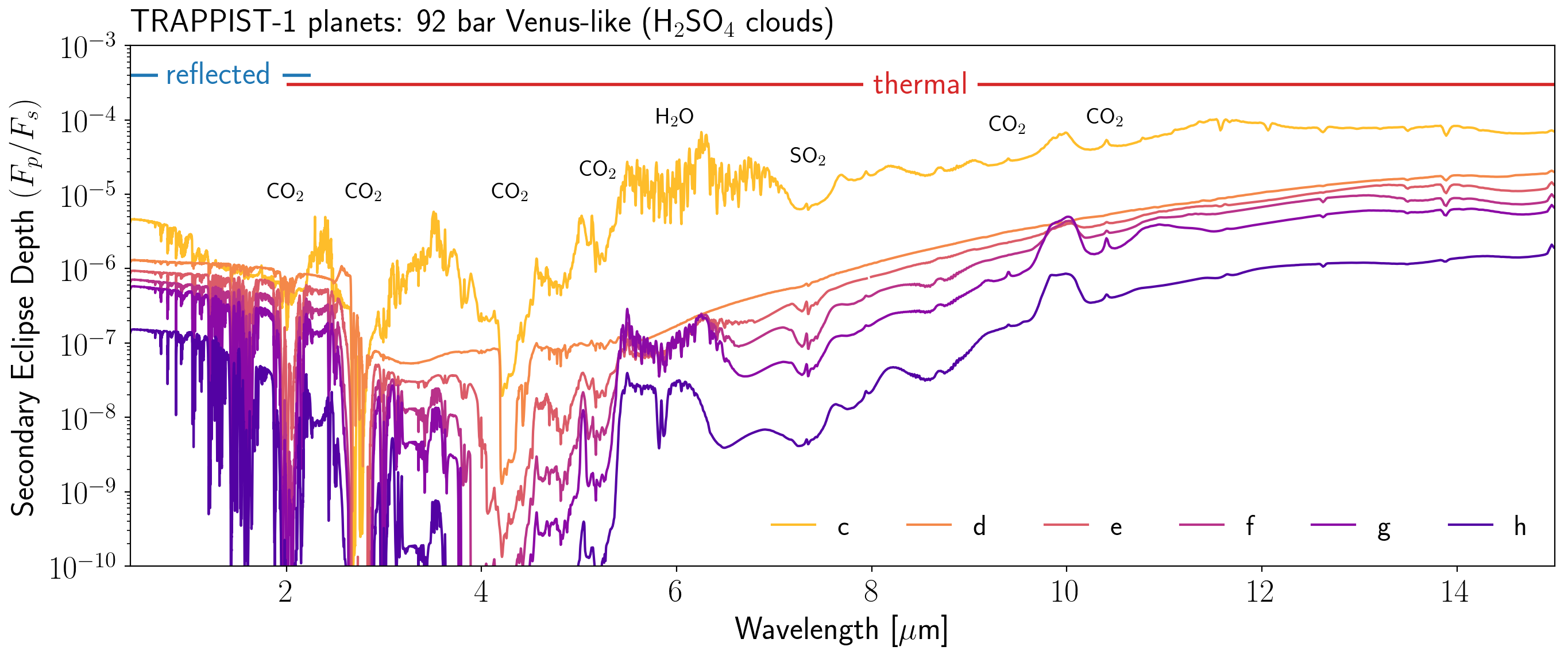}
\includegraphics[width=0.98\textwidth]{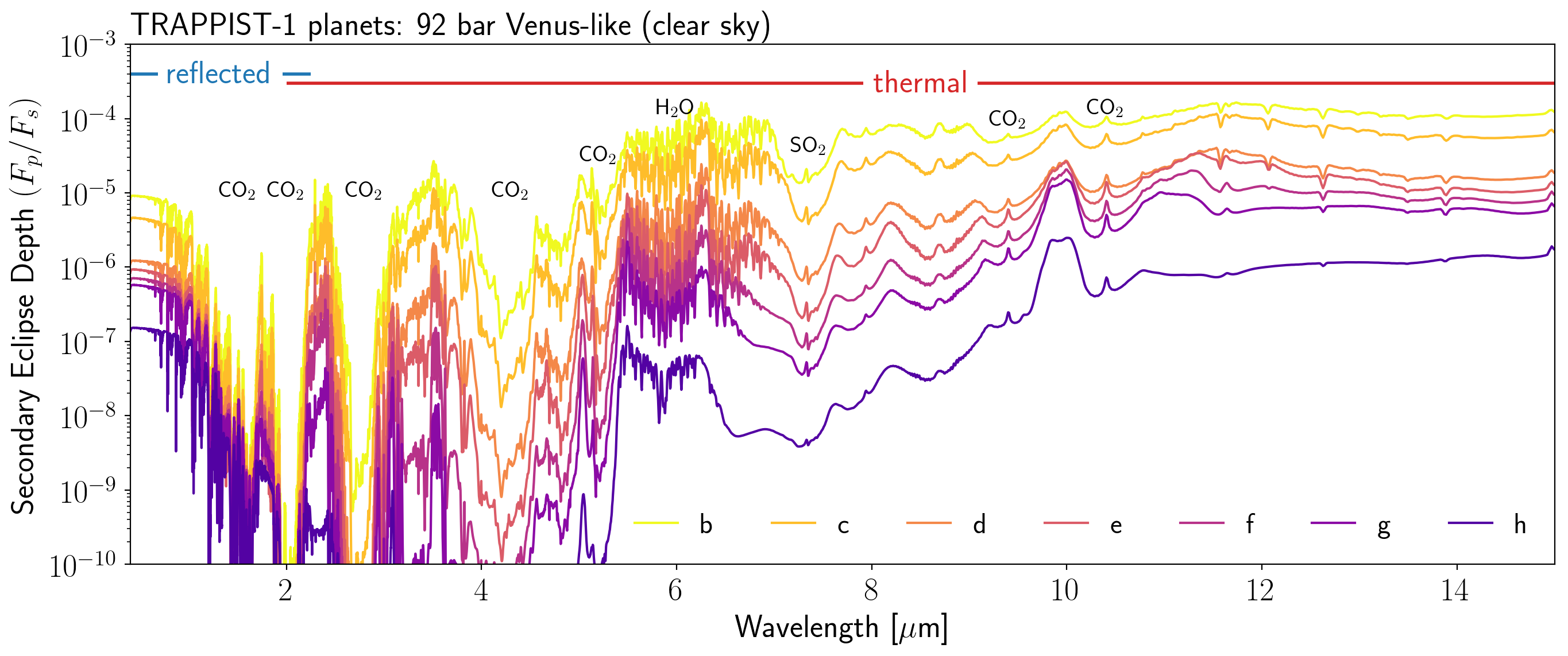}
\caption{Synthetic secondary eclipse spectra of the TRAPPIST-1 planets assuming they possess either thick Venus-like atmospheres with \ce{H2SO4} clouds (top panel) or no clouds (bottom panel). Reflected stellar flux dominates at short wavelengths, while thermal emission dominates at long wavelengths.} 
\label{fig:emission_spectra}
\end{figure*}

Figure \ref{fig:emission_spectra} shows simulated secondary eclipse spectra for our Venus-like TRAPPIST-1 models. At short wavelengths each eclipse spectrum is dominated by reflected light, while at long wavelengths they are dominated by thermal emission. Both radiative source functions decrease with semi-major axis, making eclipse spectroscopy of temperate and cool planets require at least an order of magnitude higher precision observations. Note that thick clouds can mute the thermal emission spectrum features, as best exemplified by TRAPPIST-1 d in our models. The aforementioned 6 $\mu$m spectral window shows enhanced flux of thermal radiation, particularly in the clear sky models, extending nearly 100 ppm above the thermal continuum for TRAPPIST-1 b. 

\section{Discussion} \label{sec:discussion}

%\subsection{Accessing the lower atmospheres of Venus-like exoplanets}

%Figures \ref{fig:transit_pressure}, \ref{fig:transit_probed}, and \ref{fig:emission_pressure} demonstrate the sheer difficulty intrinsic to probing the lower atmospheres of Venus-like worlds. Without even considering the uncertainty that would propagate from observation through to the inference of a maximum pressure, we see that information from the hot lower atmospheres does not exit the atmosphere. 

%However, in some cases the thermal continuum of an exo-Venus secondary eclipse spectrum may emerge from and bear evidence of a hotter, higher pressure, lower atmosphere. These atmospheric windows to lower in the atmosphere may offer the key observable necessary to break the degeneracy between cloud-top and solid surface. In particular, the 6 $\mu$m thermal window is optimally located for observations with MIRI LRS. 

%Although transparent windows may reveal lower atmospheres, they are only feasible for hot planets with high equilibrium temperatures. This is because secondary eclipse depths in the mid-IR are measures of the emitted planet thermal flux, which decay strongly with planet temperature. As a result, it is likely that only TRAPPIST-1 b and c will have strong enough thermal emission to detect the thermal emission windows \citep{Lustig-Yaeger2019, Koll2019}. 

We used the TRAPPIST-1 planets to demonstrate the difficulty intrinsic to identifying and studying the lower atmospheres of Venus-like exoplanets with transmission and emission spectroscopy, which is applicable to near-term efforts with JWST. In the case of true Venus analog exoplanets with sulfuric acid clouds and 92 bar surface pressures, transmission spectroscopy will only be sensitive to pressures between $10^2$ - $10^4$ Pa (0.001 - 0.1 bar). Although the tenuous above-cloud atmosphere could still be detected for all of the TRAPPIST-1 planets with JWST \citep{Lustig-Yaeger2019}, inferring the presence of the lower atmosphere---which perhaps best defines the very nature of Venus---will be a significant challenge, likely exceeding the scope of transmission spectroscopy. 
In the following discussion we will present observational approaches that may best constrain lower atmospheres for Venus-like exoplanets (\S\ref{discussion:lower_atmospheres}). We then discuss how when lower atmospheres cannot be observationally constrained, there exists a potential ambiguity between cloud-top and solid surface (\S\ref{discussion:cloud_surfaces}) that may pose clouds as a false positive for atmospheric erosion (\S\ref{discussion:false_positive}). We will finish with a discussion of possible strategies and opportunities to mitigate these challenges (\S\ref{discussion:mitigation}). 

\subsection{Accessing the lower atmospheres of Venus-like exoplanets} \label{discussion:lower_atmospheres}

Although emission spectroscopy is able to probe deeper than transmission, optically thick clouds may still impede lower atmosphere studies. Additionally, the NIR transparent windows that are used to probe down to the surface of Venus for spatially-resolved remote-sensing studies in our Solar System, are out of observational reach for exoplanets in secondary eclipse because dayside reflected light will overwhelm photons emerging from the lower atmosphere (see Fig. \ref{fig:emission_spectra}), analogous to how the illuminated crescent of Venus must be spatially avoided when observing nightside thermal windows \citep{Meadows1996}. Instead, thermal emission measurements that are sensitive to the lower atmosphere must push to longer wavelengths where the reflected stellar SED is naturally dimmed in the Rayleigh-Jeans tail. In particular, there is a 6 $\mu$m opacity window that is optimally located for observations with MIRI LRS. However, this MIR window does not probe within an order of magnitude of the surface pressure and falls short of the surface temperature by over $200$ K in our TRAPPIST-1 models. Additionally, water vapor has the potential to close the 6 $\mu$m opacity window, so atmospheres with more atmospheric water than the \citet{Lincowski2018} Venus-like models may not have this observable window into the lower atmosphere. 

Promising observational approaches have been proposed to efficiently identify the presence, or lack, of hot terrestrial atmospheres using photometry, and in some cases these methods may immediately favor the existence of thick atmospheres. Thermal phase curves with large day-night contrasts can rule out thick atmospheres that would otherwise redistribute heat to the nightside \citep{Seager2009, Selsis2011, Kreidberg2016, Koll2016, Kreidberg2019}, while an offset hot spot from the substellar point could favor a thick atmosphere \citep{Demory2016}. 
Similarly, secondary eclipse photometry could indicate a low dayside temperature due to atmospheric heat redistribution \citep{Koll2019} or a high bond albedo due to subsolar clouds \citep{Mansfield2019}. 
Although promising for warm to hot planets ($T_{eq} = 300-880$ K), constraints from secondary eclipses, and thermal studies in general, become quickly infeasible with JWST for cooler planets into and beyond the HZ \citep{Lustig-Yaeger2019, Koll2019}. For these cooler planets, transmission spectroscopy is especially favorable because the strength of spectral features scales with the planet's atmospheric scale height, $H = k T / \mu g \propto T_{eq}$, rather than the planet's thermal emission which scales much more strongly with temperature for temperate planets not in the Rayleigh-Jeans limit\footnote{For instance, the blackbody flux scales approximately as $T^8$ near 300 K and at 15 $\mu$m (for $< 1\%$ errors incurred by a Taylor series expansion of the exponential).} \citep{Winn2010}, as shown for the TRAPPIST-1 planets in Fig. \ref{fig:emission_spectra}. 

\subsection{An ambiguity between cloud-top and solid surface} 
\label{discussion:cloud_surfaces}

%It is for the temperate and cooler planets where a lack of evidence from thermal emission may leave transmission spectra vulnerable to misinterpretation via an ambiguity between cloud-top and solid surface. 
However, because of a lack of thermal emission data possible for the temperate and cooler planets, the interpretation of their transmission spectra is paramount in the era of JWST, but it may be complicated by an ambiguity between cloud-top and solid surface.

For any single exoplanet the presence of aerosols may be quite difficult to diagnose with transmission spectroscopy. In principle, scattering slopes and/or absorption features from aerosols may be used to break the cloud-surface degeneracy. However, high S/N observations would be needed to detect these features as they are 10-20 ppm in Venus-like TRAPPIST-1 models \citep{Lincowski2018}, which is much smaller than any of the potentially detectable spectral features with JWST \citep{Lustig-Yaeger2019}. As a result, this cloud-top--solid-surface ambiguity is more likely to emerge for spectra with low S/N either due to prohibitively long exposure times or observing strategies that seek a large sample of spectra at low to moderate S/N for statistical comparative planetology \citep[e.g.][]{Bean2017, Checlair2019}.  

\subsection{Clouds as a statistical false positive for atmospheric loss} 
\label{discussion:false_positive}

Across a population of exoplanets---either within a single planetary system, as in the case of TRAPPIST-1, or for an ensemble of planets from many systems---inferred trends in cloud-top pressure with stellar irradiation (as seen in Figure \ref{fig:transit_probed}) for similar composition atmospheres may erroneously appear as a surface pressure trend due to atmospheric loss processes. 
Specifically, the left panel of Figure \ref{fig:transit_probed} clearly shows that higher pressures are probed for cloudy exo-Venuses with larger semi-major axes. Without our prior knowledge on the inclusion of clouds in our models, and under the veil of the cloud-surface ambiguity, these trends could readily appear as trends in surface pressure. That is, \textit{are we seeing thicker atmospheres as stellar irradiation decreases, or lower cloud decks, or both?} 

This potential statistical false positive may be particularly nefarious because atmospheric loss is predicted to play a major role in sculpting the atmospheres of small rocky planets orbiting late M dwarfs. Models indicate that the TRAPPIST-1, and similar close-in, planets may have had their atmospheres completely eroded by x-ray and extreme ultraviolet radiation \citep[XUV;][]{Airapetian2017, Garcia-Sage2017, Roettenbacher2017, Zahnle2017, Dong2018, Fleming2019}, although sufficient volatile outgassing could help maintain atmospheres \citep{Bolmont2017, Garcia-Sage2017, Bourrier2017b}. 
Furthermore, \citet{Dong2018} found that the outer TRAPPIST-1 planets are capable of retaining their atmospheres over billions of years, while the inner planets may not be able to. Thus, observing a trend of thin atmospheres for the inner TRAPPIST-1 planets to thick atmospheres for the outer TRAPPIST-1 planets may appear consistent with a ``cosmic shoreline''---an empirical division between planets with and without atmospheres based on the relationship between total incident stellar radiation and planetary escape velocities \citep{Zahnle2013, Zahnle2017}. Testing the cosmic shoreline hypothesis on exoplanet data will require a statistical comparative planetology approach, as outlined in \citet{Bean2017} and \citet{Checlair2019}, but care must be taken to understand and mitigate degenerate exoplanet population trends.  

\subsection{Mitigation strategies \& opportunities} 
\label{discussion:mitigation}

Ambiguous trends in the maximum pressure seen across a population of planetary transmission spectra can also be used to implicate clouds and potentially expose their composition. 
First, \citet{Lincowski2018} found that sulfuric acid clouds did not condense in the Venus-like model atmospheres of TRAPPIST-1 b, and so our analysis did not include a cloudy TRAPPIST-1 b. However, as noted by \citet{Lincowski2019}, detecting the atmosphere of the innermost planet in multi-planet systems could strongly increase the likelihood that similar size planets at longer orbits have atmospheres, because the loss of volatiles due to escape over a planet's history is expected to decrease with increasing distance from the star. 
%Second, constraints on the composition of possible clouds could be placed by modeling the atmospheres below the continuum surface self-consistently with the modest constraints on the atmospheric composition retrieved from the transmission spectrum above the continuum surface. 
Second, by understanding what physical and chemical conditions may produce continuum pressure trends in terrestrial transmission spectra, it may be possible to rule out false positives scenarios, in the same way that false positive biosignatures may be identified and mitigated using additional environmental context from the atmosphere and stellar environment \citep{Meadows2017, Meadows2018b, Catling2018}. 
If distinct population trends seen with stellar insolation are consistent with predictions from cloud condensation modeling, then clouds could potentially be revealed by statistical characterization where they were unidentifiable in any single planet spectrum. This is a terrestrial exoplanet analog to cloud condensation trends observed in brown dwarf atmospheres across the L/T transition \citep[e.g.][]{Ackerman2001, Morley2012}.  

Furthermore, the distinctly opposing trends that we found between clear and cloudy atmospheres as a function of semi-major axis (see Figure \ref{fig:transit_probed}) highlights the potential to use the transmission spectrum continuum pressure to group similar populations of terrestrial exoplanets. The maximum pressure probed in clear atmospheres is set by the critical refraction pressure which scales with the angular size of the star as seen from the planet \citep{Betremieux2014, Misra2014a}, and prevents access to higher pressures at larger semi-major axes (for planets orbiting similar sized stars, e.g. late M dwarfs). 
%Contrasting this trend with that seen for sulfuric acid clouds may allow for the two populations to be distinguished. 
Conversely, sulfuric acid cloud condensation allows access to higher pressures at larger semi-major axes. These contrasting trends could be detected by retrieving cloud top or reference pressures for multiple planets within the same system and may enable thick clear and cloudy atmospheres to be distinguished.  

These distinguishing characteristics extend into the terrestrial domain the concepts presented in \citet{Sing2016} of an observable distinction between clear and cloudy atmospheres in an ensemble of exoplanet spectra. \citet{Sing2016} found that the strength of water absorption features in the spectra of hot Jupiter exoplanets is correlated with cloud and haze scattering slopes in the spectra, indicating that clouds/hazes may be obscuring the water column, rather than seeing an intrinsic trend in water vapor abundance. 
%Similarly, we have explored how the strength of \ce{CO2} absorption features that should be detectable with JWST is fundamentally limited in Venus-like exoplanets by how \ce{H2SO4} clouds control the spectral continuum. These clouds may be quite difficult to diagnose in transmission and emission spectra, but their presence in a set of planetary spectra may be revealed by the behaviour of the cloud-top pressure as a function of stellar insolation. 
Similarly, we have explored how the strength of gaseous absorption features relative to the spectral continuum could potentially be used to discriminate between populations of thin, thick/clear, and thick/cloudy atmospheres, even if the transmission spectra are individually difficult to diagnose. 

The transition between planets with and without atmospheres and the transition between terrestrial and gaseous planets are two bookends of the high mean molecular weight, terrestrial atmosphere regime. 
The emerging paucity of planets with radii ${\sim} 1.6$ R$_\oplus$ orbiting Sun-like stars \citep{Rogers2015, Fulton2017} likely constrains the presence of terrestrial atmospheres on the large planet boundary. 
JWST will offer a first opportunity to investigate this boundary on the small planet end, as we continue to explore the effects of atmospheric escape \citep[e.g.][]{Lehmer2017}, and attempt to map the cosmic shoreline. 

\section{Conclusions} \label{sec:conclusion}  

% Can we detect the lower atmospheres of Venus-like exoplanets? Honestly, not really. 
The lower atmospheres of Venus-like exoplanets may elude our characterization efforts with JWST due to the presence of sulfuric acid clouds, which both dictate remote studies of Venus and constitute a potential terrestrial exoplanet analog to the high altitude clouds and hazes that currently limit the charaterization of gaseous exoplanets with transmission spectroscopy. 
For hot exo-Venuses, MIR opacity windows observed during secondary eclipse may offer glimpses of thermal emission from the atmosphere just below the clouds, potentially allowing for high-pressure, greenhouse-heated lower atmospheres to be directly inferred. 

However, for temperate to cold exo-Venuses observed with transmission spectroscopy, a sulfuric acid cloud deck may appear indistinguishable from a solid surface at low to moderate S/N as both cause the spectral continuum to be flat. In these cases, Venus-like atmospheres should still be detectable via \ce{CO2} absorption features, but appear like tenuous low pressure atmospheres due to a lack of observational constraint from the lower atmosphere. 
For Venus-like atmospheric models of the TRAPPIST-1 planets, we demonstrated that the sulfuric acid clouds drop in altitude to higher pressures with semi-major axis. This effect has the potential to be misinterpreted as a trend of increasing surface pressure with decreasing stellar insolation and may appear suspiciously consistent with atmospheric escape. 
Looking ahead, the prospect of different populations of terrestrial exoplanets---cloudy exo-Venuses and stellar windswept worlds---presenting similar observables motivates the need for additional climate, photochemical, cloud formation, and atmospheric escape modeling to uncover observable characteristics that effectively discriminate between different populations of exoplanets, and observing strategies tailored to test these hypotheses.   

 %% Acknowledgements %%
\acknowledgments

\

This work was supported by NASA's NExSS Virtual Planetary Laboratory funded by the NASA Astrobiology Program under grant 80NSSC18K0829. This work made use of the advanced computational, storage, and networking infrastructure provided by the Hyak supercomputer system at the University of Washington. 

\software{Matplotlib \citep{Hunter2007}, Numpy \citep{Walt2011}, LBLABC \citep{Meadows1996}, SMART \citep{Meadows1996}}

%% Bibliography %%
\bibliography{bib}

%% End Doc
\end{document}